\documentclass[aps,prl,amsmath,amssymb,amsfont,showpacs,superscriptaddress,reprint]{revtex4-1}

\usepackage{amsmath}
\usepackage[usenames,dvips]{color}
\usepackage{graphicx}

\begin{document}

\title{Linearly controlled arrangement of $^{13}$C isotopes in single-wall carbon nanotubes}

\author{J. Koltai}
\affiliation{Department of Biological Physics, E\"{o}tv\"{o}s University, P\'{a}zm\'{a}ny P\'{e}ter s\'{e}t\'{a}ny 1/A, H-1117 Budapest, Hungary}
\author{H. Kuzmany}
\affiliation{Universit\"{a}t Wien, Fakult\"{a}t f\"{u}r Physik, Strudlhofgasse 4, 1090 Wien, Austria}
\author{T. Pichler}
\affiliation{Universit\"{a}t Wien, Fakult\"{a}t f\"{u}r Physik, Strudlhofgasse 4, 1090 Wien, Austria}
\author{F. Simon}
\affiliation{Universit\"{a}t Wien, Fakult\"{a}t f\"{u}r Physik, Strudlhofgasse 4, 1090 Wien, Austria}
\affiliation{Department of Physics, Budapest University of Technology and Economics and MTA-BME Lend\"{u}let Spintronics Research Group (PROSPIN), P.O. Box 91, H-1521 Budapest, Hungary}


\begin{abstract}
	The growth of single wall carbon nanotubes (SWCNT) inside host SWCNTs remains a compelling alternative to the conventional catalyst induced growth processes. It not only provides a catalyst free process but the ability to control the constituents of the inner tube if appropriate starting molecules are used. We report herein the growth of inner SWCNTs from $^{13}$C labeled toluene and natural carbon C$_{60}$. The latter molecule is essentially a stopper which acts to retain the smaller toluene. The Raman spectrum of the inner nanotubes is anomalous as it contains a highly isotope shifted "tail", which cannot be explained by assuming a homogeneous distribution of the isotopes. {\color{black}Semi-empirical} calculations of the Raman modes indicate that this unsual effect is explicable if small clusters of $^{13}$C are assumed. This indicates the absence of carbon diffusion during the inner tube growth. When combined with appropriate molecular recognition, this may enable a molecular engineering of the atomic and isotope composition of the inner tubes.
\end{abstract}

\maketitle

\section{Introduction}

The growth of carbon nanotubes from carbonaceous materials, which are encapsulated inside host SWCNTs, remains a compelling catalyst free synthesis method of SWCNTs. While the growth was originally discovered from encapsulated fullerenes (peapods) \cite{SmithNAT} under intensive electron beam irradiation and heating \cite{LuzziCPL2000}, it was later shown that the inner tube can proceed from virtually any carbon containing materials \cite{SimonCPL2006} including small solvent molecules such as e.g. benzene or toluene, azafullerenes \cite{SimonCAR2006}, or coronene \cite{KamarasSmall}. This synthesis method of SWCNTs have the advantage of allowing a diameter control depending on the diameter of the outer tube and that a catalyst free synthesis is performed, which leads to ultra clean inner tubes \cite{PfeifferPRL2003}. Natural disadvantages of the inner tube growth are the hindered ability to remove the inner tubes from the inside in a non-invasive manner, and the lack of control over the possible inner-outer 
tube chirality pairs, whose presence complicates the Raman analysis \cite{PfeifferPRB2005b}. 

A possible next step to explore the inner tube synthesis from various carbon sources is the combination of several starting components, e.g. the combination of co-encapsulated fullerenes and small organic molecules, which could be used e.g. for the growth of heteroatom containing inner tubes or for their isotope labeling. In principle, the various carbon source molecules would encapsulate in a random fashion. However if some sort of a molecular recognition was present, it would be a possibility to control the arrangement of the various components. In addition, it is also required that little carbon atom diffusion takes place during the inner tube growth in order to fully exploit the molecular recognition. It was reported previously \cite{ZolyomiPRB2007} that the carbon diffusion is limited along the inner tube axis: fullerenes of natural carbon and $^{13}$C were co-encapsulated and a Raman analysis of the inner tube modes showed a larger than expected inhomogeneity of the $^{13}$C isotopes on the resulting 
inner tubes. A logical continuation of this effort is to co-encapsulate a $^{13}$C isotope labeled small organic molecule (benzene or toluene) with fullerenes inside SWCNTs and to study the vibrational modes of the resulting inner tubes.

We report the synthesis and Raman characterization of single-wall carbon nanotubes which are grown inside host nanotubes from fullerenes and other small organic molecules, benzene and toluene. When the latter molecules are $^{13}$C isotope labeled, unexpected changes in the Raman spectra are observed: rather than downshifting in a uniform manner (which is expected for homogeneous doping), a tail develops on the small Raman shift side of the Raman modes. This indicates a clustering of the $^{13}$C isotopes. This is supported by first principles calculations, where similar features can only be reproduced when a significant clustering of the isotopes is present. This effect is probably related to the clustered nature of the $^{13}$C isotopes on the organic rings and it suggests that no carbon diffusion takes place during the inner tube growth.

\section{Experimental}

The starting SWCNT sample was obtained by the arc-discharge method and it was identical to samples as in previous studies \cite{SimonCPL2006} with a mean diameter of $1.4\,\rm nm$ and a Gaussian distribution varuiance of 0.1 nm. This diameter is suitable for the growth of inner tubes as it can optimally contain the filled in fullerenes. Annealing in air of the SWCNT sample for $0.5\,\rm h$ at $450^{\circ}{\rm C}$ opens the nanotubes. Commercial fullerenes (Hoechst, Super Gold Grade ${\rm C}_{60}$, purity $99.9\%$) were used along with and natural (Sigma) and $^{13}{\rm C}$ enriched benzene and toluene (Eurisotop, France). We note that for toluene, only the benzyl ring was enriched while the methyl group was of natural carbon. Co-encapsulation of the fullerenes and the benzene or toluene proceeds by sonicating the SWCNTs for 2 hours in the corresponding solvent: ${\rm C}_{60}$ solution of $1\,\rm mg/ml$. This method is known to result in a clathrate structure where the smaller benzene or toluene molecules 
occupy half-half of the available inner volume in an alternating fashion \cite{SimonCPL2006,ZolyomiJPCC2014}. The resulting material was filtered to obtain nanotube \textit{bucky-papers} and it was rinsed with the corresponding non-enriched solvent (benzene or toluene) to remove any non-encapsulated excess fullerenes from the outside of the nanotubes. This step was followed by the final filtering and drying of the \textit{bucky-paper} samples under a fume hood. The samples were annealed in dynamic vacuum at $1250\,^{\circ}{\rm C}$ for 1 hour. This process is known to yield high quality double-wall carbon nanotubes \cite{SimonPRB2005,SimonCPL2006}. In the following, we denote double-wall carbon nanotubes grown from ${\rm C}_{60}$ and toluene both containing natural carbon as $^{12}{\rm DWCNT}$. $^{12/13}{\rm DWCNT}$ denotes double wall carbon nanotubes where the inner wall is made of co-encapsulated ${\rm C}_{60}$ and $^{13}$C benzene and ring-enriched toluene.

Raman spectroscopy was performed with a Dilor xy triple monochromator spectrometer with an excitation lines of an Ar-Kr gas discharge laser. We report data with the $514.5\,\rm nm$ line of the Ar ion only; the energetic 2D Raman line of the inner tubes is in resonance with this excitation \cite{Pfeiffer2005PRB} i.e. its observation is most convenient.

\section{Raman spectroscopy results}

\begin{figure}[h]
\includegraphics[width=\linewidth]{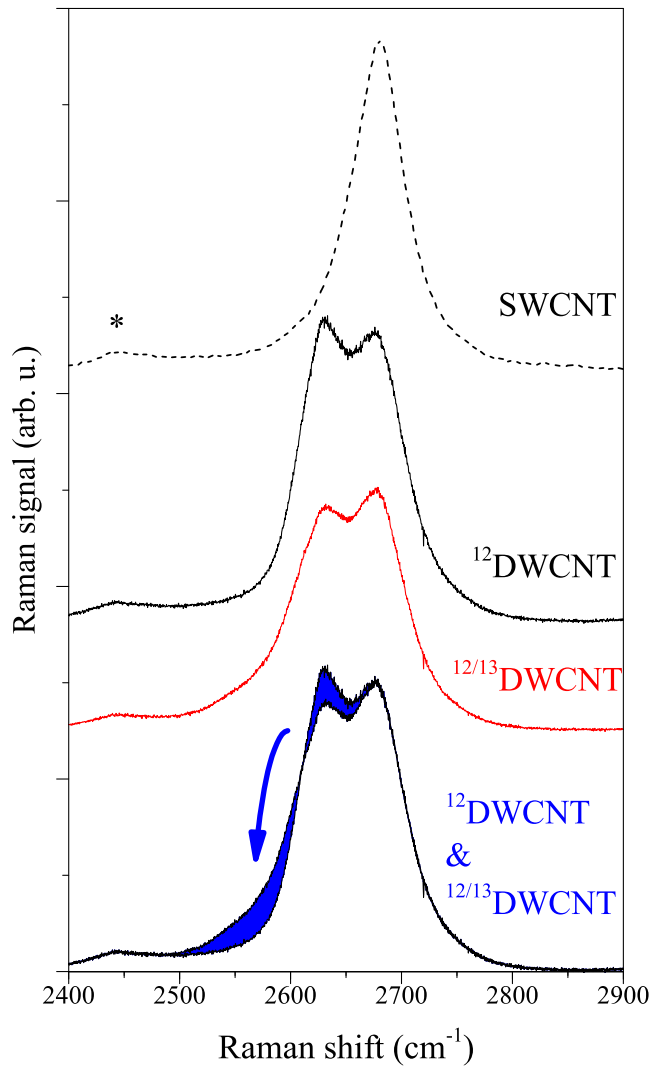}
\caption{Raman spectra of the toluene+C$_{60}$ based DWCNT samples around the 2D Raman line. A corresponding Raman spectrum on the starting SWCNT sample is shown for comparison (dashed curve). The $^{12}$DWCNT and $^{12^13}$DWCNT data are shown with an offset first in the middle. The two spectra are shown on one another at the bottom in order to highlight the spectral weight which is downshifted (blue filled area, indicated by an arrow). Asterisk indicates a small Raman line which is present in the pristine sample already.}
\label{KB_proc_Fig1}
\end{figure}

We show the Raman spectra of reference SWCNT, $^{12}$DWCNT and $^{12/13}$DWCNT samples around the 2D Raman line spectral range. The latter sample was based on a mixture of ring-labelled $^{13}$C$_6$H$_6$-CH$3$ and natural carbon toluene with a ratio of 84:16. This means that the nominal $^{13}$C carbon content in the toluene was 72 \%. The DWCNT samples are characterized by the emergence of the lower frequency 2D Raman line which corresponds to the inner tubes \cite{Pfeiffer2005PRB}. In fact, the inner tube 2D Raman line is twice as strong as the present one for C$_{60}$ based inner tubes. This is explained by the fact that toluene and benzene are relatively large compared to their nominal carbon content as compared to the fullerenes \cite{SimonCPL2006}. It means that they use a significant amount of the available volume while contributing to less carbon atoms to the inner tube growth. 

The Raman spectra of $^{12}$DWCNT and $^{12/13}$DWCNT shows a striking difference which is clearest from the bottommost comparison in Fig. \ref{KB_proc_Fig1} (indicated by a blue arrow in the figure): a sizeable amount of spectral weight is shifted from around the peak of the inner tube 2D Raman line toward lower Raman shifts which forms a "tail". This is a surprising observation as a homogeneous downshift of the Raman line is expected from a naive consideration of isotope labeling, rather than the formation of a low Raman frequency "tail". Previously, the growth of inner tubes was investigated from $^{13}$C enriched fullerenes and therein a uniform downshift was observed with a mean downshift corresponding to the formula \cite{SimonPRL2005,ZolyomiPRB2007}:

\begin{gather}
\frac{f_0}{f}=\sqrt{\frac{12.011+c\cdot 13}{12.011}}
\label{eq:downshift}
\end{gather}

\noindent where $f$ and $f_0$ are the Raman shifts with and without $^{13}{\rm C}$ doping, $12.011\,\rm g/mol$ is the molar mass of natural carbon and it reflects the $1.1\%$ abundance of $^{13}{\rm C}$ in natural carbon and $c$ is the $^{13}{\rm C}$ concentration. We note that previously a similar, anomalous development of a low Raman frequency tail was observed in isotope labeled benzene based inner tubes in Ref. \cite{KoltaiJPCC2016}. 

\begin{figure}[h]
\includegraphics[width=\linewidth]{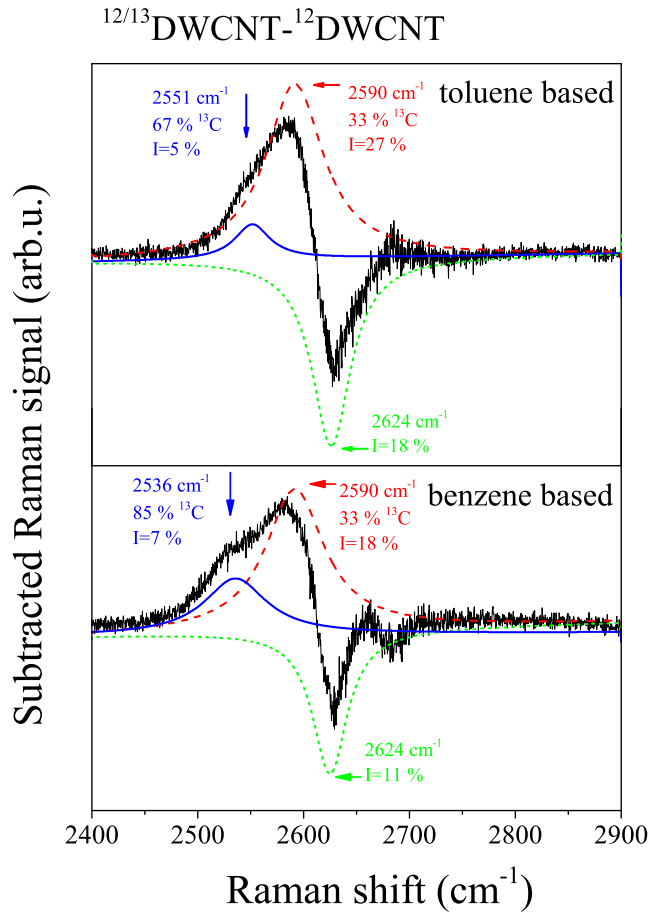}
\caption{The spectrum obtained after subtracting the 2D Raman line of the $^{12}{\rm DWCNT}$ sample from the $^{12/13}{\rm DWCNT}$. A deconvolution into several components is also shown. The Raman shifts, the nominal $^{13}{\rm C}$ enrichment level and the intensity of the particular component with respect to the inner tube 2D Raman line are shown. For comparison we show the same kind of data for benzene based $^{12/13}{\rm DWCNT}$ sample after Ref. \cite{KoltaiJPCC2016}.}
\label{KB_proc_Fig2}
\end{figure}

We show in Fig. \ref{KB_proc_Fig2}. the spectrum obtained after subtracting the 2D Raman line of the $^{12}{\rm DWCNT}$ sample from the $^{12/13}{\rm DWCNT}$. We also show for comparison the same kind of data on $^{13}$C labeled benzene based DWCNTs from Ref. \cite{KoltaiJPCC2016}. The deconvolution of the subtracted spectrum into Lorentzian components reveal three components: a significantly downshifted, low intensity line, a moderately downshifted stronger signal, and a line at the position of the unenriched inner tubes with a negative intensity. The latter corresponds to the spectral weight which is missing in the subtracted spectrum, i.e. it is the spectral weight which is downshifted for the isotope labeled sample in agreement with Fig. \ref{KB_proc_Fig1}. The significantly downshifted component corresponds to a $^{13}$C isotope enrichment of $67\,\%$. This is smaller than the enrichment found for the same line in the benzene based DWCNT in Ref. \cite{KoltaiJPCC2016} which may be due to the lower ($72\,\%$)  
$^{13}$C isotope enrichment of carbon of toluene as compared to $99\,\%$ in benzene. An interesting observation is that the other downshifted Raman line is found at the same position for both kinds of samples. 

It is tempting to associate the significantly downshifted Raman line to a localized cluster or island of  $^{13}$C isotopes, which assumption is studied further below. The analysis of the Raman intensity shows for both kinds of samples that no Raman spectral weight conservation applies.

\section{Theoretical modelling and discussion}
{\color{black}
We calculated the first order Raman spectrum with the semi-empirical PM3 method as implemented in the Gaussian09 package \cite{g09}. 
We neglected the outer tube and the inner tube was considered as a molecule: a hydrogen-terminated piece of (5,5) armchair type SWCNT consisting of 600 carbon and 20 hydrogen atoms. The structure was first relaxed with the {\em opt=tight} option, then we obtained the force constants and the polarizability derivatives. We compared the Raman spectra of various small molecules (methane, benzene, ${\rm C}_{60}$ fullerene) calculated with the PM3 method to the first principles based (DFT/B3LYP) results.
In general, the frequencies obtained by the semi-empirical method are not very accurate, which could be fixed by rescaling the force constants to fit the experimental (or first principles) value,
but we do not bother the absolute position of the calculated Raman peaks, because we always used the same force constant matrix and only changed the masses in the dynamical matrix according to the isotope distribution.
However, the Raman intensities calculated with the PM3 method reproduced the results of the more demanding DFT/B3LYP method very well. 
Using the {\em freqchk} utility of Gaussian09 \cite{g09} for different distribution of isotope masses the Raman intensities were then evaluated for 2000 random configurations. Two kinds of isotope distribution were considered i) the homogeneous distribution where 60 carbon atoms were selected and replaced individually, and ii) ring arrangement where 10 complete rings of 6 carbon atoms were substituted by $^{13}{\rm C}$ isotopes -- both resulting in a nominal $10\,\%$ isotope enrichment.

In Figs. \ref{KB_proc_Fig3} and \ref{KB_proc_Fig4} the data of each vibrational mode of every configurations are presented in the G band region for the homogeneous and the clustered distribution, respectively. The colors (and sizes) of symbols correspond to the Raman intensity, the $x$-axis is the usual Raman shift and the $y$-axis is the weight of the $^{13}{\rm C}$ substituted carbon atoms movements of the vibration mode. 
The overall left-top right-bottom trend confirms the naive expectation, that the more dominant the motion of $^{13}{\rm C}$ atoms in a normal mode are, the more significant its redshift is.
The horizontal solid (black) line marks the position of the isotope-shifted G peak according to Eq. \ref{eq:downshift} with $c=10\,\%$. In the homogeneous case the shift can be well described with this simple formula, while in the clustered arrangement the shift is clearly larger and could be fitted with an effective enrichment of $c=16\%$. The vertical solid (black) line indicates the weight of the $^{13}$C substituted carbon atoms ($1/10$) in the normal modes regarding the nominal $10\%$ enrichment. For homogeneous distribution of the isotopes the center of the peaks lies very precisely on this line -- to no surprise. However, for the ring configurations there is a convincing deviation from the simple $1/10$ value. This is telling us, that if the $^{13}$ substituted carbon atoms are clustered (e.g. in a ring), their presence can give more weight to the normal mode and also to the Raman intensity of the specific normal mode. Therefore they give more weight to the averaged Raman spectrum \cite{KoltaiJPCC2016}. This effect might lie behind the anomalous development of a low Raman frequency. It is also worth to note, that this is specific for the G mode, we did not find a similar behavior for the Radial Breathing Mode.

Our theoretical finding supports the experimental observation, that upon  clusterization the downshift of the Raman peaks can be higher than the real substitution ratio. 
However, we did not find as significant downshifts as measured. This might be due to the specific character of the 2D Raman line \cite{ThomsenPRL2000}. A similar analysis for the 2D Raman line would be computationally extremely demanding, since there is an additional integration over k-space for the 2D Raman line.

\begin{figure}[h]
\includegraphics[width=\linewidth]{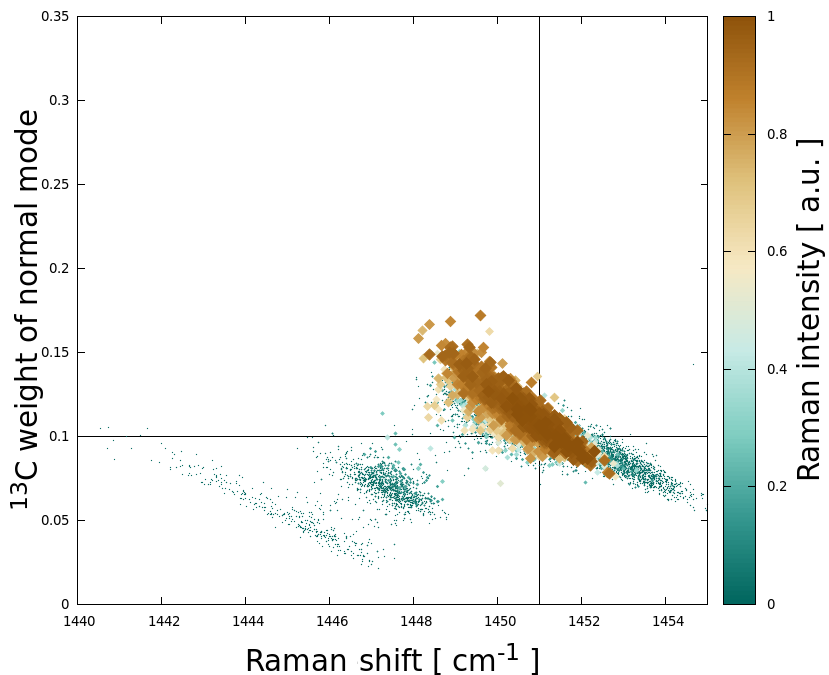}
\caption{Scatterplot showing the calculated Raman intensities for the homogeneous distribution of $^{13}{\rm C}$ isotopes.}
\label{KB_proc_Fig3}
\end{figure}

\begin{figure}[h]
\includegraphics[width=\linewidth]{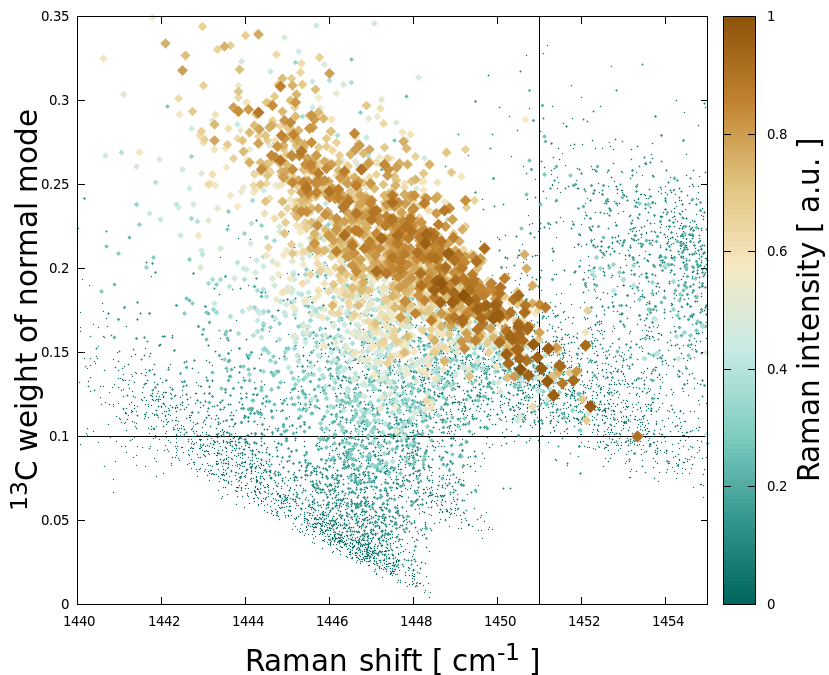}
\caption{Scatterplot showing the calculated Raman intensities for the ring arrangement of $^{13}{\rm C}$ isotopes.}
\label{KB_proc_Fig4}
\end{figure}

}

\section{Conclusions}
In conclusion, we presented the synthesis and Raman characterization of carbon nanotubes grown from $^{13}$C isotope labeled organic solvent (benzene and toluene) inside host outer tubes. Raman spectroscopy analysis indicates that the $^{13}$C isotopes are non uniformly distributed on the inner tube walls. This indicates that little or no carbon diffusion takes places during the inner tube growth. The analysis is supported by {\color{black}semi-empirical} calculations of the vibrational modes for clustered $^{13}$C isotope rich configurations. The material with $^{13}$C isotope rich clusters may find application as local nuclear spin labels or in quantum information storage.

\section{Acknowledgement}
The Hungarian National Research, Development and Innovation Office (NKFIH) Grants Nr. K108676, K115608, and K119442 are acknowledged for support.

\providecommand{\WileyBibTextsc}{}
\let\textsc\WileyBibTextsc
\providecommand{\othercit}{}
\providecommand{\jr}[1]{#1}
\providecommand{\etal}{~et~al.}

\end{document}